\documentclass[12pt]{article}
\usepackage{aas}
\usepackage{times}
\usepackage{epsfig}
\begin{document}
\centerline{\bf  Continuous heating of a giant X-ray flare on Algol}

\vskip 0.5cm
\centerline{J.H.M.M. Schmitt$^*$ \& F. Favata$^{**}$}
\vskip 0.25cm
$^{*}$: Universit\"at Hamburg, Hamburger Sternwarte, 21029 Hamburg, 
Gojenbergsweg 112
\hfil\break
$^{**}$: ESA-ESTEC, Space Science Department, ESTEC , Postbus 299, NL-2200 AG
Noordwijk
\hfil\break
\vskip 0.25cm
{\bf 
Giant flares can release large amounts of energy
within a few days$^{1-7}$: X-ray emission alone can be up 
to ten percent of the star's bolometric luminosity.
These flares exceed the luminosities of the largest
solar flares by many orders of magnitude, which suggests that the
underlying physical mechanisms supplying the energy are
different from those on the Sun.  Magnetic
coupling between the components in a binary system or between a young star 
and an accretion disk has been 
proposed$^{3,7-9}$ as a prerequisite for giant flares.
Here we report X-ray observations of a giant flare
on Algol B, a giant star in an eclipsing 
binary system.  We observed a total X-ray eclipse of the flare, which
demonstrates that the plasma was confined to Algol B, and reached
a maximum height of 0.6 stellar radii above its surface.  The flare occurred
around the south pole of Algol B, and energy must have been 
released continously throughout its life. We conclude that
a specific extrastellar environment is not required for the presence of 
a flare, and that the processes at work are therefore similar to those
on the Sun.}

X-ray and radio observations have produced copious evidence for 
magnetic-field related activity on nearly all cool stars
with outer surface convection zones$^{10,11}$.
Magnetic activity on other stars is usually interpreted by
analogy with the Sun, viewing the observed
stellar emissions as coming from scaled-up versions
of directly observable solar features.
The physical validity of this approach is debatable,
particularly for the observed extremes of stellar activity, which exceed
the corresponding solar emissions by at least five orders of magnitude.  

Magnetic reconnection is important in the theory of solar flares$^{12}$, 
yet the application of the theory to solar observations produces
ambiguous evidence for the occurrence of reconnection$^{13,14}$. 
Reconnection is a local phenomenon; the difficulty lies in 
explaining how energy can be released swiftly and
coherently from rather small volumes.
This becomes even more difficult when discussing the 
energetics of giant stellar flares.  At peak, such flares can
release up to a few percent of the quiescent bolometric luminosity, and their
overall energy release equals the cumulative output of a few hours of
quiescent luminosity; there must be, therefore, be an efficient way
to first store, and then release, vast amounts of energy.
Extreme levels of activity are observed in close binary 
systems and in very young stars, where the magnetic field topology 
is likely to differ from that of the Sun.
In a close binary system, magnetic field lines
may connect the two components, enabling plasma to be magnetically
confined over a much larger interbinary volume; similarly, in 
a star surrounded by an accretion disk magnetic 
field lines originating in the photosphere may thread the disk and magnetic 
stresses may be built up by shearing motions.  
It has therefore been proposed
that the specific environment of such systems is a prerequisite for the 
occurrence of giant flares$^{3,7-9}$. 
\hfil\break
\begin{table*}
\caption[ ]{\label{tab1}System Parameters for Algol ($=\beta Per$)}
\begin{flushleft}
\begin{tabular}{| r | r | r | r |}
\hline
Parameter& Primary &  Secondary  & System \cr 
\hline
Mass (g) & 7.54\ 10$^{33}$ & 1.64\ 10$^{33}$ & \cr
\hline
Radius (cm) & 2.15\ 10$^{11}$ & 2.29\ 10$^{11}$ &\cr
\hline
Spectral type        & B8V             & K2III & \cr
\hline
Luminosity (erg/sec) &  5.95\ 10$^{35}$ & 2.44\ 10$^{34}$ & \cr
\hline
Rotation period (days)      & 2.8673   & 2.8673 & \cr
\hline
Orbital period (days)        &       &    &   2.8673 \cr
\hline
Distance  (cm)             &       &    &   1.02\ 10$^{12}$ \cr
\hline
\end{tabular}
\end{flushleft}
\end{table*}

In eclipsing binaries, size information can be obtained from light curves.
The eclipsing binary Algol is among the nearest and X-ray brightest 
eclipsing systems; the relevant system parameters$^{15}$
 are given in Table 1.
Frequent X-ray flares have been reported on 
Algol$^{4,16,17}$ and
the absence of an X-ray eclipse at optical secondary minimum has been 
interpreted as evidence for a corona with a scale height of more 
than a stellar radius$^{18}$.
A 2.9 day X-ray observation of Algol, covering the whole binary orbit, was
carried out with the Italian BeppoSAX satellite$^{19}$ .  Here we 
discuss only the medium energy concentrator spectrometer
(MECS) light curve in the energy band 
between 1.6 - 10~keV.  In Fig. 1 we show the phase folded BeppoSAX MECS
light curve of Algol which is dominated by a huge flare lasting almost
through the entire observation.  The observed peak luminosity is 
3$\times$ 10$^{32}$ erg/sec (at least 1.2 \% of the late-type
component's quiescent luminosity), and
integration over all of the available BeppoSAX light 
curves yields a total observed X-ray
energy release $E_{X-ray,tot} \approx 1.5\times 10^{37}$ 
erg in the 0.1 - 10~keV band; thus 
the flare is at least as energetic as
the largest flares observed on RS CVn systems, T Tauri
and proto-stars$^{1-7}$.

At phase $\phi \ = $ 1.5 a count rate minimum is observed (cf., Fig. 1).  The 
only plausible interpretation of this light curve
feature is an X-ray eclipse of the flaring plasma by the
early-type primary.  
Residual X-ray flux persists at minimum, but a number of indications
suggest that flux to come from Algol's quiescent (unocculted) corona
and not from the flare:  At $\phi \ = $ 1.5 the light curve is 
straight while
it decreases and increases before and after that phase; at minimum the
observed count rate is close to the count rates before flare onset,
and also the observed count rate at minimum is very similar to the
mean ROSAT all-sky survey count rate.  It is thus reasonable to assume that
the eclipse of the X-ray flare is total.  In order to validate this 
assumption, we attempted to determine
the appropriate quiescent emission level at phase $\phi \ = $ 1.5.  All
values between 0.4 and 1.5 cts/sec can be chosen and need to be compared
to the observed minimum rate of 1.08 cts/sec (cf. Fig. 2, upper panel).  
We thus estimate that at least 90\% of the flare emission was occulted;
as a temperature increase was detected only
when the count rate being above 1.1 cts/sec,
we assume in the following that the eclipse of the flare was total. 

Three important features of the BeppoSAX
X-ray light curve are evident: 
(1) a total eclipse of the X-ray flare by the X-ray
dark early-type primary;  
(2) a symmterical pattern of eclipse ingress and egress with a
shallow light curve decrease of $\sim$ 20 \% 
followed by a steep decrease of the remaining 80 \% on ingress and vice
reversed on egress;  
(3) no evidence for any occultation of the flare plasma by the 
late-type secondary (no rotational modulation; note that 
because of synchronous rotation
two different stellar hemispheres are in view
at $\phi \sim $ 1.5 during minimum and at $\phi \sim $ 1.0 when the flare 
erupts).
The occurrence of an eclipse at  $\phi \ = $ 1.5
indicates that the flare is associated with the secondary;
the absence of rotational modulation then implies that the
flare plasma must be located near one of the poles.  
The eclipse totality
restricts the maximal height above the surface
to $d_{AB} \ sin(i) = R_{*}$, where $d_{AB}$ is the distance between
the binary components Algol A and Algol B, $i$ is the angle between orbit
plane normal and line of sight, and $R_{*}$ is the radius is the star,
considerably less than a stellar radius.
A three-dimensional view of the admissible flare locations is 
displayed in Fig. 3. The flare undoubtedly occurred above the south 
pole of Algol B and not in the interbinary region.

For the maximally available volume in Fig. 3 we derive a value of 
V$_{max}\ = 1.5\  \times 10^{33}$ cm$^{3}$.  
Clearly, the actual flare is likely to occupy only some fraction
$f_{max}$ of V$_{max}$.  
>From spectral analysis$^{20}$ 
the total emission measure at the flare peak, $EM = {n^2} {V}$ = 1.33 
$\times 10^{55}$cm$^{-3}$, and the temperature T = 1 $\times 10^8$~K
can be determined. 
Hence, the
flare plasma density $n$ = 9.4$\times 10^{10}/f_{max}^{1/2}$ 
cm$^{-3}$, and the thermal pressure is 2.6$\times 10^{3}/f_{max}^{1/2}$ 
dyn cm$^{-2}$.  Given these values the radiative cooling time$^{17}$
$\tau _{rad}$ = $\frac {3 k T } {n P(T)}$, where $k$ is Boltzmann's
constant, and the cooling function
$P(T) = 10^{-24.73} T^{0.25}$) (in CGS units), becomes
$\tau _{rad} = 23600 \times f_{max}^{1/2} $ sec.  This is 
much smaller than the observed light curve decay time scale $\tau _{dec} 
\approx 60 000$ sec regardless of $f_{max}$.  It seems logical to  conclude 
that continued heating through
continued energy release must be present throughout the flare.
Of course, the flare plasma may cool conductively.  Determining
the conductive cooling time scale$^{17}$, we find
$\tau _{cond} = 2530 \times f_{max}^{-1/2}$ sec by setting the adopted
length scale
$L = V_{max}^{1/3}$.  This value of $\tau _{cond}$
is smaller than $\tau _{rad}$ for large filling
factors;  however, $\tau _{cond}$ is sensitively dependent on $L$, 
and we are thus unable to prove that
$\tau _{cond} < \tau _{rad}$  holds.

Small heights above the surface and continued heating are
characteristic properties of ``two-ribbon flares'' on the
Sun$^{21}$.
The energy for continued heating
is derived from the reconnection of (non-potential) magnetic fields 
of opposing polarity.  
If all of the released energy $E_{released}$
is derived from the magnetic energy in 
the volume $V_{max}$, the minimally required magnetic field 
strength $B_{min}$ can be computed by equating
$\frac {B_{min}^2\ V_{max}} {8\ \pi}\ = \ \frac {E_{released}} {f_{X-ray}}$,
where f$_{X-ray}$ denotes the observed fraction of the released energy.
We find $ B = 500/\sqrt{f_{X-ray}}$ G to be the (non-potential)
pre-reconnection magnetic field that needs to be annihilated 
to meet the energy requirements.  
If, in addition,
the hot X-ray emitting plasma generated in the reconnection
process must be magnetically confined, field strengths of at least
B = 256$/f_{max}^{1/4}$ G after reconnection are required.  

Our finding that the Algol flare occurred above the south pole of 
Algol B, fits well to the presence ``polar spots'' 
in many rapidly rotating active binary systems$^{22,23}$.  
While no
magnetic fields measurements nor Doppler images have been made of Algol B,
it is reasonable to assume the presence of such polar spots on Algol B.
Photospheric magnetic field strengths$^{24}$ on late-type stars 
are typically in the range 1000 - 2000 G, and these values should also apply
to Algol B.  These values are clearly a strict upper limit to the coronal
magnetic field strengths.

A solar-like reconnection scenario thus appears to be most natural
to explain all the observations of the giant flare on Algol.  
For physical consistency, magnetic fields of 500~G - 1000~G,
much larger than in the corona of the Sun$^{21}$,  must 
be present in the corona of Algol B 
at heights of up to half a stellar 
radius, extending over a volume of $> 10^{33}$ cm$^3$ and reconnected in 
the course of the flare.  The flaring plasma is associated only 
with the late-type star, and the binary nature of Algol is irrelevant, at 
least for this giant flare.  
The flare does, however, also display several
features not observed in solar flares.
Solar flares are 
never observed in polar regions; instead they are confined to the
active region belt at lower heliographic latitudes.  
The thermal plasma in the giant flare on Algol is dominated 
by temperatures of $1 \times $ 10$^8$ K; such temperatures are
occasionally observed in solar flares$^{25,26}$ as ``super hot plasmas'', but
they contain only small fractions of the overall emission measure.
The derived minimal plasma density is similar to the plasma 
densities of many solar flares$^{27}$, which would require 
a volume close to $V_{max}$ to satisfy the observed energy budget.
Higher densities with correspondingly smaller volumes can, of course, not
be excluded; however, much higher densities quickly lead to 
implausibly large thermal pressures.  Thus,
the real challenge to theory is the construction of
physically consistent
reconnection based flare models with the observed
stellar parameters and clarify the role
of giant flares for the mass and angular momentum loss of active stars.

\begin{figure*}[htbp]
  \begin{center}
    \leavevmode \epsfig{file=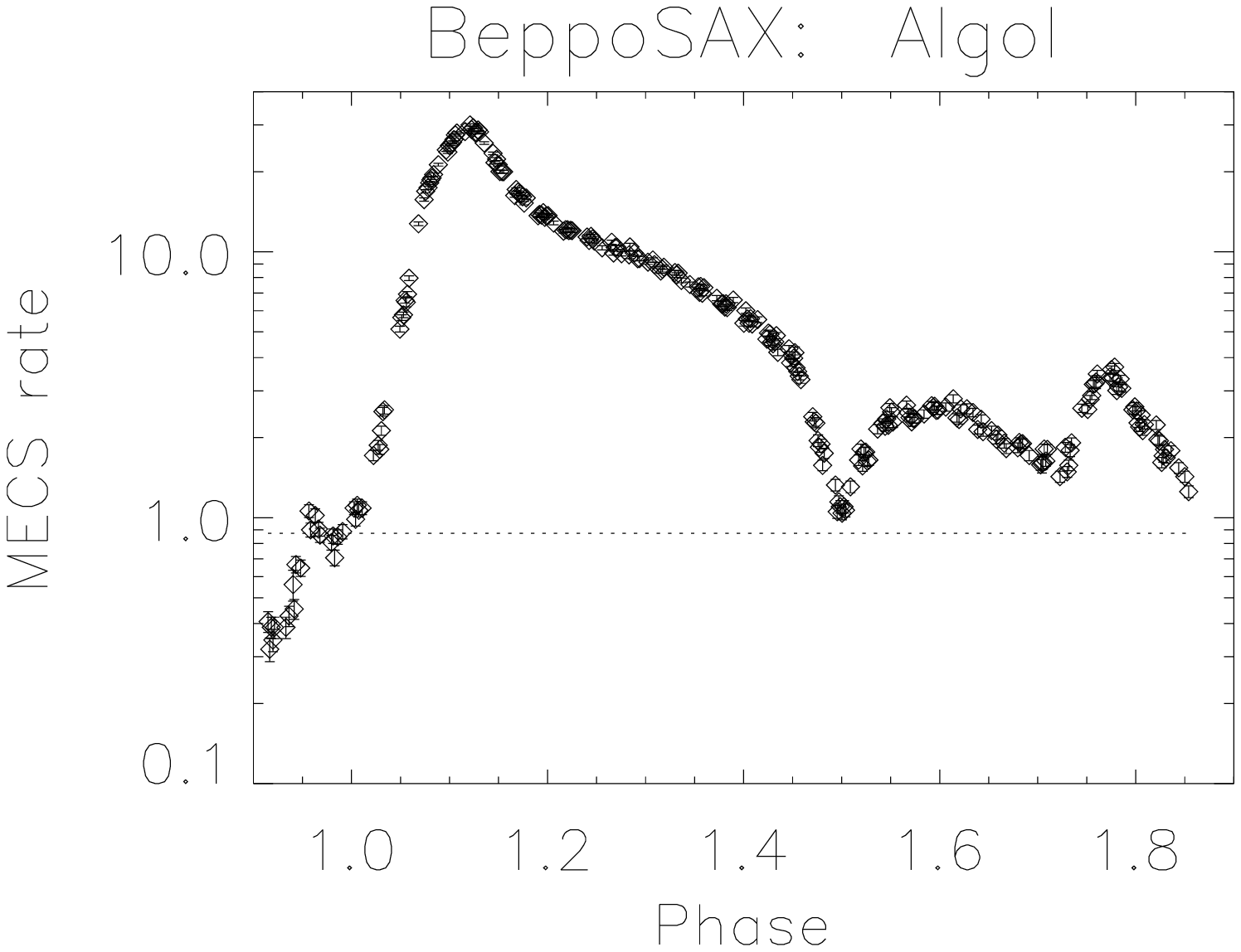,}
    \caption{X-ray light curve of Algol measured with the SAX MECS2-MECS3
detectors in the 1.6 - 10  keV band 
between Aug 30, 1997 3:04 UT and Sep 1, 1997, 20:32 UT
with phase $\phi$ calculated from
the ephemeris JD = 2445739.003 + 2.8673285 $\times$ E 
(E integer) for the times 
of primary minimum$^{15}$; a hundredth of a 
phase corresponds to 41.3 minutes.
The phasing is such that for $\phi _{E} = E\ +\ 0.5$ the
X-ray dark early type primary is in front of the X-ray emitting 
late type secondary.
Note the huge flare starting at phase $\phi \ \sim $ 1.0 with a
rise time of about 8.3 hours and peaking at $\phi \ \sim $ 1.12.
A rapid initial decay until at $\phi \ \sim $ 1.25 is followed
by an exponential decay.  A clear eclipse of the flaring plasma is seen at
$\phi \ \sim $ 1.5, when the early type primary is in front of the
late-type secondary.   The dotted line indicates an estimate of the
quiescent out-of-flare rate, extrapolated from the observed ROSAT
all-sky survey count rate.}
  \end{center}
\end{figure*}

\begin{figure*}[htbp]
  \begin{center}
    \leavevmode \epsfig{file=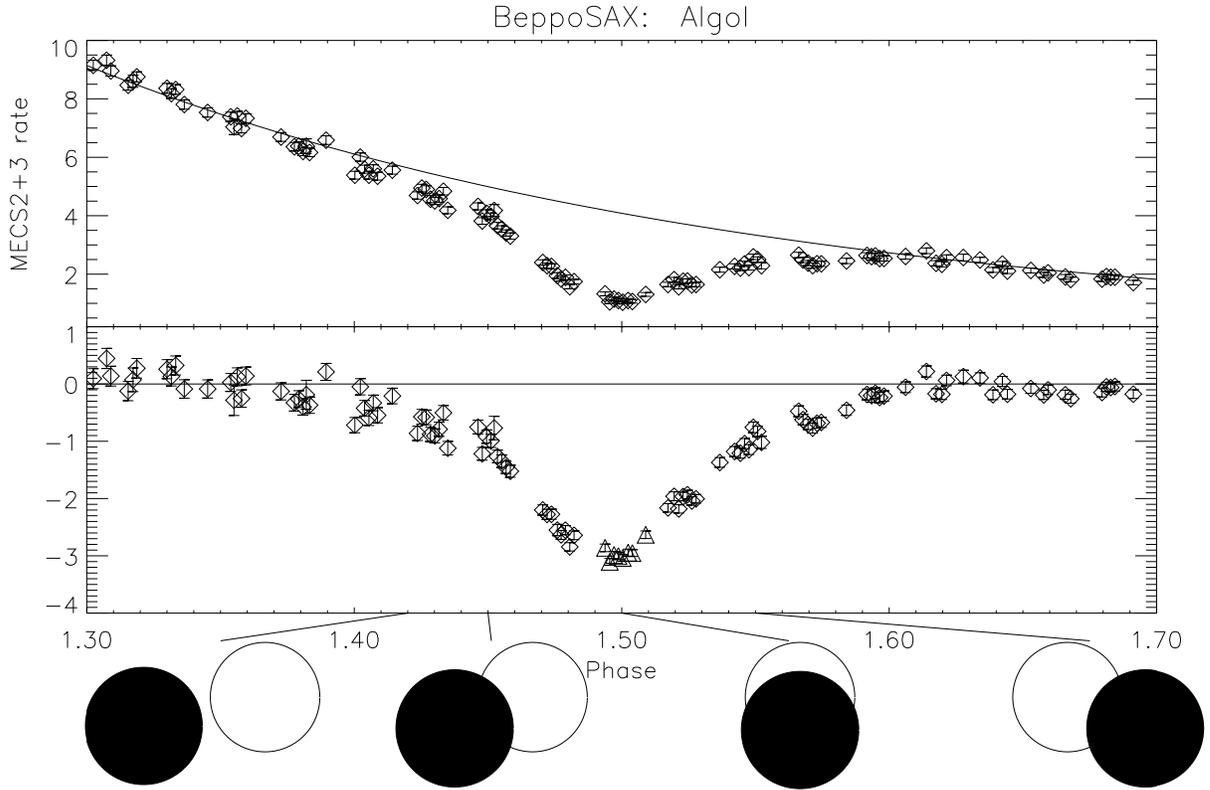,}
    \caption{ a) Upper panel: The MECS count 
rate vs. phase in the interval 1.3 - 1.7;
the solid line represents an exponential fit to the pre- and
post eclipse light curve.
b) Lower panel:  Count rate vs. phase in the interval 1.3 - 1.7
with exponential decay (shown in panel a) removed; the zero line is
shown. 
The flare eclipse starts at $\phi \sim $ 1.39 with a somewhat shallow
decay, whence at $\phi \sim $ 1.451 a sharp decay starts (first contact).  
Totality begins at $\phi \sim $ 1.488 (second contact) 
and ends at $\phi \sim $  1.505 (third contact).  The 
light curve increases quickly until  $\phi \sim $  1.545 (fourth contact)
and then more slowly until $\phi \sim $  1.605, when the flare eclipse is 
over.  Below panel (b) the viewing geometry is shown at phases
$\phi = $ 1.42, 1.45, 1.50, and 155; the filled circle represents
Algol A as it moves across Algol B.
}
  \end{center}
\end{figure*}

\begin{figure*}[htbp]
  \begin{center}
    \leavevmode \epsfig{file=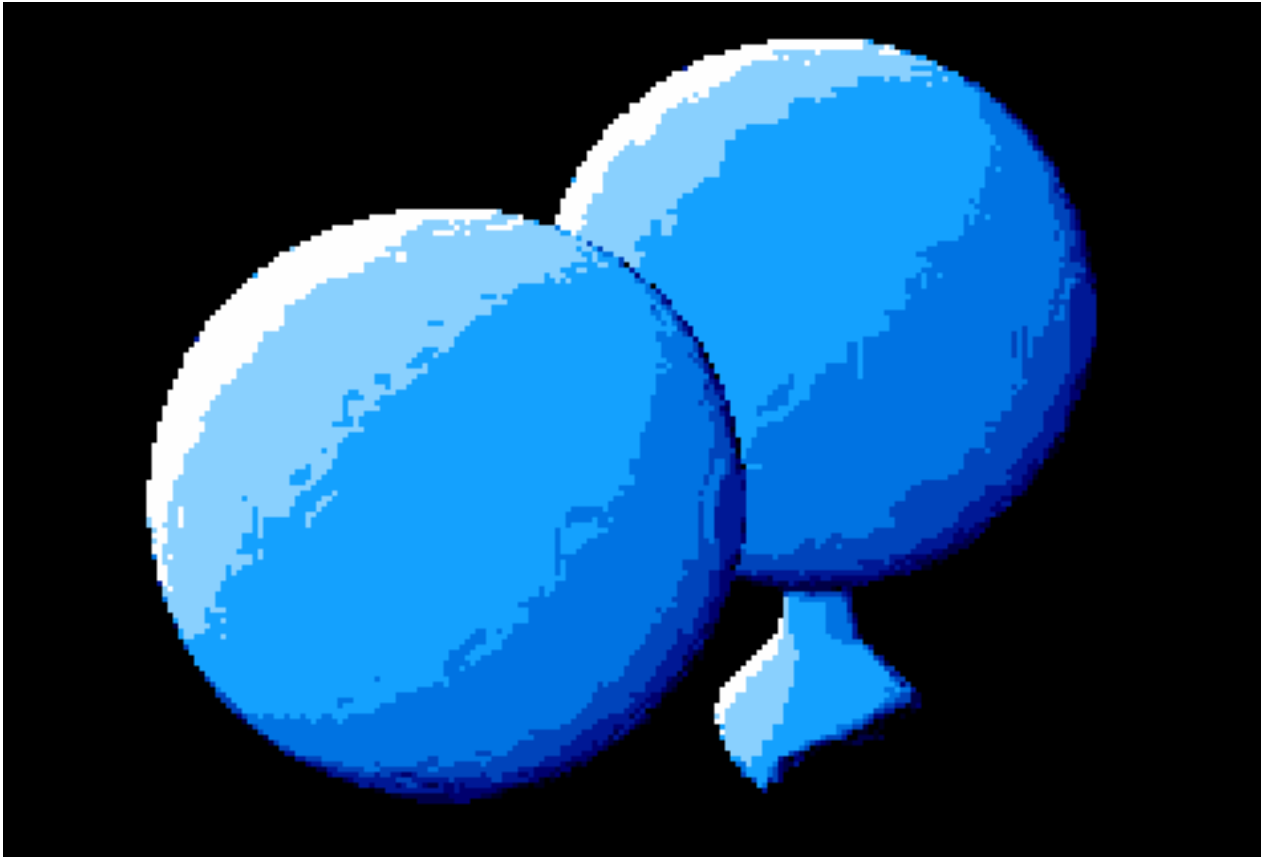,}
    \caption{3D-Visualisation of the Algol system including the
maximum volume containing the flare. Shown are
the two binary components Algol A (foreground) and Algol B (background) 
as seen from our line of sight at phase $\phi$ = 1.45.  The
extension below the south  pole of Algol B
is the computed locus of all points satisfying the following requirements:
(1) they must be 
totally eclipsed by Algol A between second and third contact;
(2) they must be not eclipsed by  Algol A and not occulted 
by  Algol B itself
before first and after fourth contact;  (3)
they must be visible between first and second but not between third
and fourth contact or but not both; (4)  
assuming that
the flaring plasma originates from the surface of Algol B, along any 
radial line of sight from the surface of the  Algol B to the point under
consideration there should be no more than twenty percent
rotational modulation.  These requirements define
the hose-like structure above the south pole of
of Algol B, within which the flare must 
have occurred.  The flare is clearly associated with Algol B and did not
occur in the interbinary region.
}
  \end{center}
\end{figure*}
\end{document}